# Potential of Domain Adaptation in Machine Learning in Ecology and Hydrology to Improve Model Extrapolability

Haiyang Shi[1,2]

[1] Department of Civil and Environmental Engineering, University of Illinois at Urbana-Champaign, Urbana, IL 61801, USA

[2] The National Key Laboratory of Ecological Security and Sustainable Development in Arid Region, Chinese Academy of Sciences, Urumqi, 830011, China

Correspondence: Haiyang Shi (hiayang@illinois.edu)

**Abstract**

Due to the heterogeneity of the global distribution of ecological and hydrological ground-truth observations, machine learning models can have limited adaptability when applied to unknown locations, which is referred to as weak extrapolability. Domain adaptation techniques have been widely used in machine learning domains such as image classification, which can improve the model generalization ability by adjusting the difference or inconsistency of the domain distribution between the training and test sets. However, this approach has rarely been used explicitly in machine learning models in ecology and hydrology at the global scale, although these models have often been questioned due to geographic extrapolability issues. This paper briefly describes the shortcomings of current machine learning models of ecology and hydrology in terms of the global representativeness of the distribution of observations and the resulting limitations of the lack of extrapolability and suggests that future related modelling efforts should consider the use of domain adaptation techniques to improve extrapolability.

**Main**

In classical machine learning, it is assumed that the training and test sets are drawn from the same distribution. However, this assumption may not always hold true in real-world applications in fields like ecology and hydrology, which involve geographical variability. In the construction of big data models in ecology and hydrology, when models are applied to specific locations, there can be significant differences (dissimilarities) in the domain distributions of the test and training sets, potentially leading to decreased model performance (Shi et al., 2023). Domain Adaptation learning (Farahani et al., 2021), a type of transfer learning, aims to enhance the model's self-adaptive ability for the prediction set by adjusting the deviation between the domains of the training and test sets. Domain adaptation has been widely applied in fields like computer vision (Farahani et al., 2021), but is rarely used in big data models in ecology and hydrology. This indicates that previous models have overlooked the significant differences in domain distributions between test and training sets implied by geographical variability (including differences in climate, hydrology, underlying surfaces, etc.), highlighting an urgent need for improvements to substantially increase the models' generalizability.

Domain adaptation learning was initially proposed in the field of machine learning (Ben-David et al., 2006) and has since been widely applied in areas such as computer vision (Long et al., 2015). This approach arises when training data does not accurately reflect the distribution of test data, potentially leading to performance degradation when the trained model is applied to the test data. To address this issue, domain adaptation was introduced (Ben-David et al., 2006). In this context, the training and test sets are referred to as the source domain and target domain, respectively. Domain adaptation seeks to identify and adjust common underlying factors between the source and target domains to reduce mismatches in the feature space between the domains (Ben-David et al., 2006). Existing domain adaptation methods can be broadly categorized into shallow and deep architectures. Shallow domain adaptation methods (Gopalan et al., 2011; Pan et al., 2010) primarily use instance-based and feature-based techniques to align domain distributions. This includes minimizing the distance between domains (common distance metrics include Wasserstein distance, Kullback-Leibler divergence, etc.), and aligning the covariance of source and target domain data (correlation alignment methods) (Sun et al., 2016). On the other hand, deep domain adaptation methods (Ganin & Lempitsky, 2015; Long et al., 2015) employ neural networks. These methods often use convolutional networks, autoencoders, or adversarial networks to reduce differences between domains. Such research has demonstrated higher accuracy in a series of image classification experiments (Ganin & Lempitsky, 2015; Liang et al., 2020; Long et al., 2015) and effectively improves adaptability in situations with significant domain variations.

In the field of geospatial big data research, while some studies in remote sensing image classification (Matasci et al., 2015; Tuia et al., 2016) have begun to explore domain adaptation learning, its application in constructing big data models for hydrology and ecology, which are characterized by high geographic variability, remains limited. In many mainstream data products of hydrological and ecological geographic variables based on machine learning, all available observational data are used to build a unified prediction model, which is then directly applied on a regional or global scale (Hassani et al., 2020; Jung et al., 2020; Li et al., 2023). However, the training sets for these models are often densely distributed in economically developed regions like the United States and Europe (Figure 1), but are sparse in many underdeveloped areas and regions with minimal human footprint, such as Africa, Central Asia, and Siberia. The challenges in data sharing also mean that the availability of training set data is not maximized. As a result, current machine learning models in the fields of ecology and hydrology are more likely to learn patterns specific to regions with dense observational networks, such as the United States and Europe, and are insufficiently trained on areas with scarce observations. Therefore, the extrapolative potential of these models has not been fully recognized. This highlights a significant gap in the predictive capabilities of these models across different geographic regions, indicating a need for models that can adapt to diverse geographical conditions and data distributions.

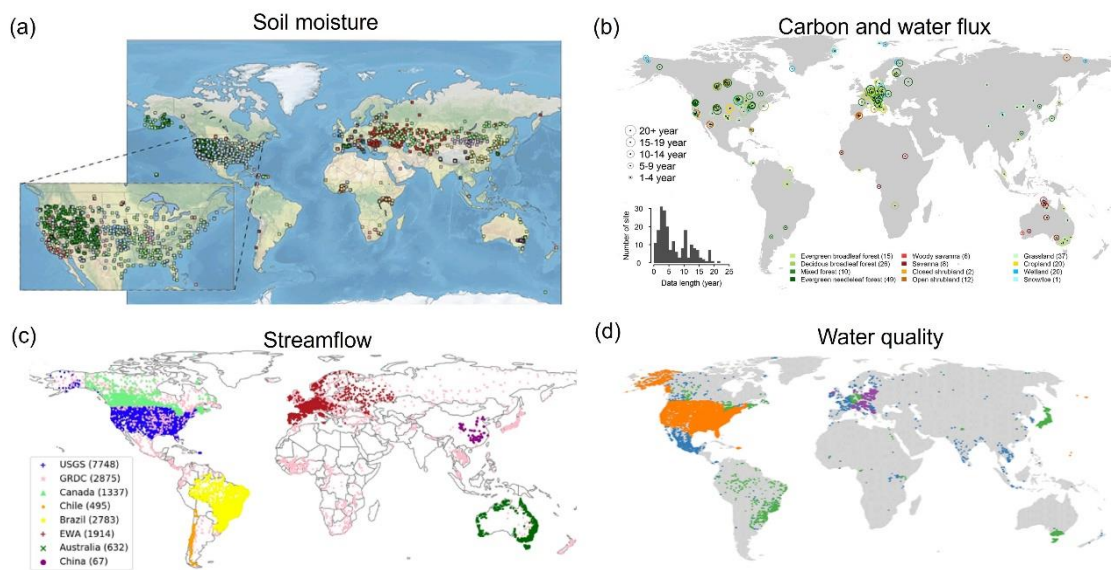

**Figure 1:** Distribution of observational data available for training global ecological and hydrological machine learning models. (a) Network of soil moisture observation stations, modified from the ref. (Dorigo et al., 2021), (b) Network of carbon and water flux stations, modified from the ref. (Pastorello et al., 2020), (c) Network of runoff gauge stations, modified from the ref. (Lin et al., 2019), (d) Network of water quality (total suspended solids) stations, modified from the ref. (Virro et al., 2021).

Many models in geospatial research employ methods like "leave-one-out cross-validation" (e.g., in the ET product of FLUXCOM (Jung et al., 2019)) or random cross-validation (such as the commonly used k-fold cross-validation (Shi et al., 2022a, 2022b)) to assess model accuracy. However, these approaches often overlook the significant impact that differences in domain distribution (or dissimilarity) between the left-out site or fold and the training set can have on the precision of predictions. These models typically don't assess and mitigate the differences between the source domain (training set) and the target domain (prediction set), leading to potentially unreliable predictions in locations where there is a substantial difference from the training data's domain. A few studies have focused on "model extrapolation" related to domain adaptability (Jung et al., 2020; Meyer & Pebesma, 2021), attempting to establish a relationship between model accuracy and the dissimilarity between the test set and the training set in leave-one-out cross-validation (Meyer & Pebesma, 2021). This approach aims to provide a geographical distribution of predictability. For instance, in ET (Evapotranspiration) simulation predictions, Shi et al. (Shi et al., 2023) compared the accuracy of models trained with all data against those trained with data from specific land cover types. They found that in some sites, models trained with all data were less accurate than those trained with specific land cover types. This suggests that quantifying the distance between the prediction set and the training set could be crucial in understanding these differences. However, these studies have not explicitly employed domain adaptation methods to optimize the domain distribution of the training set data. By doing so, they could potentially improve the adaptability of the models to different test sets, significantly enhance the accuracy of the models' predictions, and produce high-precision data on a global scale. There's a growing recognition of the need for such domain adaptation in ecological and hydrological modeling, especially considering the diverse and often sparse data distributions in these fields.

This indicates that domain adaptation techniques have not yet been fully applied in global-scale ecological and hydrological machine-learning models. If implemented, these techniques could significantly reduce inconsistencies between training and prediction sets, thereby enhancing the adaptability of models to unknown locations. While there may still be challenges in finding similar training data in extreme outlier conditions, domain adaptation offers the potential to improve adaptability for many test locations.

In summary, the application of domain adaptation in the current field of hydrological and ecological geospatial big data modelling is quite scarce, and its importance has been largely overlooked. The incorporation of domain adaptation methods in future models deserves more attention and experimentation to enhance their generalizability and extrapolation capabilities. As the field continues to evolve, recognizing and addressing these gaps can lead to more robust and accurate predictive models, especially in areas with sparse data distributions.